\documentclass[reprint, superscriptaddress, 
 amsmath,amssymb,
 aps,
]{revtex4-2}

\usepackage{graphicx}
\usepackage{dcolumn}
\usepackage{bm}

\usepackage{siunitx}
\usepackage{physics}
\usepackage{makecell}
\usepackage{subcaption}
\usepackage{bm}
\usepackage{mathtools}
\usepackage{tikz}

\newcommand{\diag}{{\tt diag}}
\newcommand{\argmin}{\mathop{\tt arg~min}\limits}

\begin{document}

\preprint{APS/123-QED}

\title{Phase retrieval via Zernike phase contrast microscopy\\with an untrained neural network}

\author{Zinan Zhou}
\affiliation{Department of Physics, Graduate School of Science, The University of Tokyo}
\author{Keiichiro Toda}
\affiliation{Institute for Photon Science and Technology, Graduate School of Science, The University of Tokyo}
\author{Rikimaru Kurata}
\affiliation{Department of Information Physics and Computing, Graduate School of Information Science and Technology, The University of Tokyo}
\author{Kohki Horie}
\affiliation{Department of Physics, Graduate School of Science, The University of Tokyo}
\author{Ryoichi Horisaki}
\affiliation{Department of Information Physics and Computing, Graduate School of Information Science and Technology, The University of Tokyo}
\author{Takuro Ideguchi}
\affiliation{Department of Physics, Graduate School of Science, The University of Tokyo}
\affiliation{Institute for Photon Science and Technology, Graduate School of Science, The University of Tokyo}
\email{ideguchi@ipst.s.u-tokyo.ac.jp}




\date{\today}

\begin{abstract}
Zernike's phase contrast microscopy (PCM) is among the most widely used techniques for observing phase objects, but it lacks quantitative nature, as it cannot directly provide phase information. Current methods for computationally extracting phase distributions from PCM images, however, rely heavily on empirical regularization parameter tuning. In this paper we extend an existing approach by employing an untrained neural network as an image prior, removing the need for manual regularization. We quantitatively demonstrate improved accuracy and robustness in phase retrieval compared to existing methods, using numerical and experimental PCM images. Our results confirm the feasibility of applying deep priors for phase retrieval in incoherent illumination setups.
\end{abstract}

\maketitle


\section{Introduction}
Phase contrast microscopy (PCM), invented by Frits Zernike in the 1930s \cite{zernike1955discovered}, has been crucial in biological and medical research, particularly in the observation of living cells in their natural state without staining. In a PCM, the light diffracted by the sample interferes with the unaltered light, forming a contrast image from which phase information about the sample can be observed. However, this imaging process is not quantitative due to its intrinsic halo and shade-off artifacts \cite{shaked2012biomedical}. Quantitative phase imaging (QPI) is desirable because it can provide more precise information about optical properties of the sample. Typically, nevertheless, QPI systems are complex and require expertise for practical usage. There is thus a need to enhance the quantitativeness of PCM while maintaining its ease to use.

One way to improve PCM is to modify its hardware setup, especially illumination and phase modulation, to reduce artifacts and improve quantitative accuracy. Maurer \textit{et al.\ }\cite{maurer_phase_2008} employed a random dot phase mask to suppress both the halo and shade-off effects. This idea was then extended by Gao \textit{et al.\ }\cite{gao_phase-shifting_2011}, where a rotatable phase plate with three phase-shifting positions was used. By phase-shifting, they achieved quantitative phase measurement, but the accuracy was limited. In spatial light interference microscopy (SLIM) \cite{wang_spatial_2011}, a similar phase-shifting technique was implemented but by an add-on module with a spatial light modulator (SLM), enabling quantitative phase imaging.

Apart from hardware modifications, purely computational methods have also been developed. These methods model the optical properties of the PCM, and phase retrieval is achieved by solving the inverse imaging problem by optimization. Yin \textit{et al.\ }\cite{yin_understanding_2012} developed a linear imaging model that allows phase retrieval within a weak phase range. Recently, we have put forward a method based on modeling the incoherent illumination, where restoration of phase within an unambiguous phase range of $\pi$ \si{\radian} has been demonstrated \cite{kurata2024single}.
However, this previous approach requires heavy manual tuning of regularization, limiting its usage in practice.

Inspired by previous works from computer science and phase retrieval \cite{jagatap2019algorithmic, wang2020phase, zhou2020diffraction, bostan2020deep, mashiko2023Scatter}, in this paper we propose using a phase retrieval algorithm with an untrained neural network (UNN) as a structural prior, which obviates the need for manual regularization terms.  ``Untrained'' means that no dataset is required for network training. To our knowledge, this is the first implementation of phase retrieval with a deep prior in an incoherent illumination configuration. We show that our strategy can be applied to a wide range of samples under PCM without adjusting hyperparameters, demonstrating its improved generalizability and practical convenience.

\section{Method}

\begin{figure}[tb]
    \centering
    \resizebox{\linewidth}{!}{
        \begin{tikzpicture}[font=\footnotesize]
            \node at (0, 0) {\includegraphics[width=.9\linewidth]{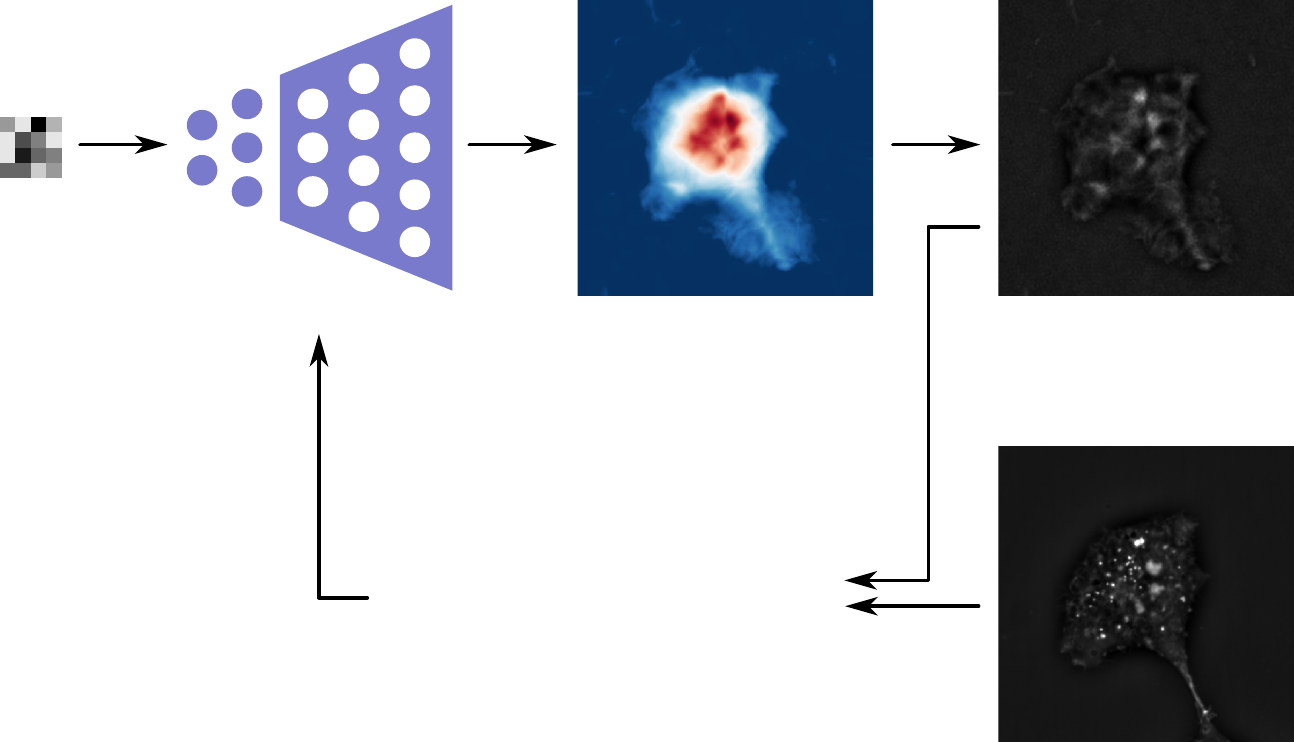}};
            \node at (-3.7, 1.8) {Fixed Tensor};
            \node at (-3.7, .9) {$\bm{B}_0$};
            \node at (-2.2, 2.5) {Untrained Network};
            \node at (-2.2, 0.5) {$\mathcal{D}\qty(\vb{W};\cdot)$};
            \node at (0.4, 2.5) {Estimated Phase};
            \node at (0.4, 0.2) {$\bm{\theta} = \mathcal{D}\qty(\vb{W};\bm{B}_0)$};
            \node at (3, 2.5) {Estimated Intensity};
            \node at (3, 0.2) {$\widehat{g}=\mathcal{H}\qty(\bm{\theta})$};
            \node at (3, -0.2) {Observed Intensity};
            \node at (3, -2.4) {$\bm{g}$};
            \node at (-0.25, -1) {Minimize};
            \node at (-0.25, -1.35) {$\norm{\bm{g} - \mathcal{H}\qty(\mathcal{D}\qty(\vb{W};\bm{B}_0))}^2$};
            \node at (-2.8, -0.5) {Update $\vb{W}$};
        \end{tikzpicture}
    }
    \caption{Schematic of the optimization. Our method aims to minimize the $\ell_2$ distance between the measured PCM images and the estimated one generated by the untrained neural network and the PCM model.}
    \label{fig:flow}
\end{figure}
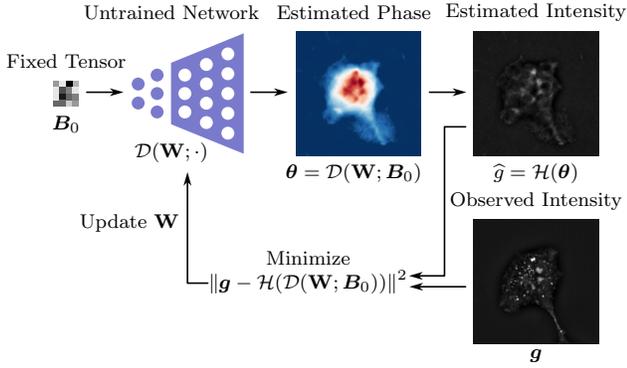

The overall flow of our proposed method is shown in Fig.~\ref{fig:flow}.
The phase image of an object is reconstructed from a single observed PCM intensity image by minimizing the error between the actual and estimated PCM intensity images. The key idea of our approach is to employ an untrained neural network as an image prior to represent the estimated phase image in the minimization problem \cite{ulyanov2018deep}. This image prior acts as a self-learned regularization, restricting the solution space. As a result, network outputs whose statistics resembles that of a natural image are favored.

With a deep network image prior~$\mathcal{D}$, a randomly initialized tensor input~$\bm{B}_0\in\mathbb{R}^{(n_0\times n_0)\times k_0}$, and network parameters~${\bf W}\coloneqq\{\bm{W}_i\in\mathbb{R}^{k_i\times k_{i+1}}\mid i = 0, \dots, 5\}$ connecting each layer, where $i$ denotes the layer index, the estimated phase image vector~$\bm{\theta}\in\mathbb{R}^{(256\times 256)\times 1}$ is represented by
\begin{equation}
\bm{\theta}=\mathcal{D}({\bf W};\bm{B}_0).
\label{eq:theta}
\end{equation}

We employ a deep decoder as the image prior $\mathcal{D}$, which is a simple image model with no convolutions and relatively few parameters \citep{heckel2018deep}. Starting from the fixed tensor $\bm{B}_0$, the tensor $\bm{B}_i\in\mathbb{R}^{(n_i\times n_i)\times k_i}$ at the next layer is computed using $\bm{W}_i\in\mathbb{R}^{k_i\times k_{i+1}}$, which is the matrix for linear combination of channels between successive layers:
\begin{equation}\label{eqn:iter}
    \bm{B}_{i+1} = {\tt ChannelNorm}\qty({\tt ReLU}\qty(\bm{U}_i\bm{B}_i\bm{W}_i))\qc i=0, \dots, 4.
\end{equation}
Here, $\bm{U}_i\in\mathbb{R}^{(n_{i+1}\times n_{i+1})\times( n_i\times n_i)}$ is the matrix performing bi-linear upsampling which doubles the canvas size $n_i$. $n_i$ is defined as 8 for $i=0$ and $n_{i+1} = 2n_i$ for $i \geq 0$. $k_i$ is the number of channels, set to 128 for $0\leq i\leq 5$.
``{\tt ReLU}'' is chosen to be the activation function, followed by a channel normalization (``{\tt ChannelNorm}'' in \eqref{eqn:iter}).
The phase image vector $\bm{\theta}$ is formed using $\bm{B}_5$ and $\bm{W}_5\in\mathbb{R}^{{k_5\times k_6}}$, where $k_6=1$ to truncate the channel dimension to one, as follows:
\begin{equation}
    \bm{\theta} = {\tt LeakyReLU}\qty(\bm{B}_5\bm{W}_5),
\end{equation}
Unlike the original deep decoder architecture, where a sigmoid activation was applied in the final layer, we applied ``{\tt LeakyReLU}'' activation. We assume that every phase image contains a background area in its field of view, which is unoccupied by any object and hence has a phase value near zero. The purpose of employing the final {\tt leakyReLU} layer is to capture this feature.

Based on the deep image prior $\mathcal{D}$, which generates the phase image vector~$\bm{\theta}$, the optimization problem for the phase retrieval via PCM is expressed as
\begin{equation}\label{eqn:cost}
    \argmin_{{\bf W}} \norm{\bm{g} - \mathcal{H}\qty(\mathcal{D}\qty({\bf W};\bm{B}_0))}_2^2,
\end{equation}
where $\bm{g}\in\mathbb{R}^{(256\times 256)\times 1}$ is the observed PCM intensity image vector, $\mathcal{H}$ is the forward propagation model of PCM, and $\norm{\cdot}_2$ denotes the $\ell_2$ norm.

In the optimization problem shown in \eqref{eqn:cost}, the estimated PCM intensity image vector $\widehat{\bm{g}}$ is observed through the PCM model $\mathcal{H}$ by inputting the estimated phase vector $\bm{\theta}$. The PCM model in this study employs compressive propagation, with modeling and parameter settings following those of Kurata \textit{et al.\ }\cite{kurata2024single}. Compressive propagation approximates spatially partially coherent illumination by a set of random wavefronts, significantly reducing the number of coherent propagations required in inverse problems and making the computation more feasible \cite{horisaki_compressive_2022}. The forward propagation model $\mathcal{H}$ of PCM is described as
\begin{equation}
\begin{split}
     \widehat{\bm{g}}&=\mathcal{H}\qty(\bm{\theta})\\
     &= \frac{1}{M}\sum_{m=1}^M\abs{\bm{F}^{-1}\diag(\bm{p})\bm{F}\diag(\exp(\mathrm{j}\bm{\theta}))\bm{F}^{-1}\diag(\bm{c})\bm{r}_m}^2,\label{eqn:pcm}
 \end{split}
 \end{equation}
where $\bm{r}_m\in\mathbb{C}^{(256\times 256)\times 1}$ is the randomly generated wavefront vector and $m\in\{1,\dots,M\}$ is its index.
$\bm{c}\in\mathbb{R}^{(256\times 256)\times 1}$ and $\bm{p}\in\mathbb{C}^{(256\times 256)\times 1}$ represent the pupil filter vectors for the condenser annulus and the phase ring, respectively.
$\bm{F}\in\mathbb{C}^{(256\times 256)\times(256\times 256)}$ and its inverse $\bm{F}^{-1}$ denote the forward and inverse Fourier transform matrices, respectively.
``{\tt diag}'' is an operator that produces a diagonal matrix with the parenthesized vector as its diagonal elements.

The intensity image $\bm{g}$ was obtained either from an actual PCM for physical demonstration or simulated from a known ground truth phase image using the PCM model in \eqref{eqn:pcm} with 4000 ($M$) random wavefronts for numerical demonstration. The network parameters $\vb{W}$ are initialized randomly, and the target function in \eqref{eqn:cost} is optimized using the Adam optimizer \cite{kingma2014adam} at a learning rate of \num{3e-4}. For each optimization step, 200 ($M$) random wavefronts are used to compute the estimated intensity image, providing a balance between reducing statistical noise and minimizing computational burden. The number of epochs is set to 4000. Due to random initialization of ${\bf W}$ and stochasticity in $\bm{r}_m$, restoration results may fluctuate slightly over each trial.

\section{Results}
Our proposed method aims to eliminate the need for hyperparameter tuning in the regularization-based approach \cite{kurata2024single}. There, at least three hyperparameters substantially affect phase retrieval quality. The parameter $\rho$ in the alternating direction method of multipliers (ADMM) plug-and-play denoising controls the overall regularization strength \cite{chan2016plug}. The parameters $\epsilon_\text{TV}$ and $\epsilon_{\ell_1}$ are stability factors used for reweighting \cite{candes2008enhancing}, which adaptively adjust the regularization strength of the total variation (TV) to enhance smoothness and the $\ell_1$ norm to suppress background noise, respectively. We post-processed the recovered phase images $\bm {\theta}$ by segregating the object and its background and subtracting the average of the background phase value. In other words, the average background phase is set to zero.

In order to make a fair comparison, we have selected three representative setups for the regularization-based method, corresponding to high, mid, and low regularization strength, as shown in Table \ref{tab:reghyper}. The low strength combination is based on settings from previous work, whereas the high strength is determined through empirically tuning to optimize performance. There may exist combinations that lead to better results. However, it is impractical to perform an exhaustive search over all hyperparameter combinations. Moreover, the best hyperparameter combination varies with optical and morphologic properties of the object, which is precisely the issue that we aim to address. On the other hand, for our proposed method, as mentioned earlier, both the network architecture and the optimization process are fixed.

\begin{table}[tb]
\caption{Selected hyperparameter combinations in the regularization-based method.}
    \begin{ruledtabular}
    \begin{tabular}{lccc}
         & $\rho$ & $\epsilon_{\text{TV}}$ & $\epsilon_{\ell_1}$ \\
         \colrule
        High & 3000 & 200 & 100 \\
        Mid & 1000 & 50 & 50 \\
        Low & 100 & 10 & 10\\
    \end{tabular}
    \end{ruledtabular}
    \label{tab:reghyper}
\end{table}

Firstly, we prepared the following samples for numerical experiment (See Appendix \ref{sec:samples}): 1) 19 images of microbeads composed of polymethyl methacrylate (PMMA), 2) 45 images of living COS-7 cells, with well-separated cells in the field of view (FOV), 3) 92 images of living COS-7 cells, but with grouped cells touching each other in the FOV, 4) 15 images of phase-only resolution targets. The ground truth phase distributions of the samples~1)--3) were measured using a homemade digital holography (DH) system based on the common-path broadband diffraction phase microscopy technique \cite{bhaduri2014diffraction}. This system contains a commercial microscope (Olympus IX73) equipped with a $40 \times 0.6$ NA objective (LUCPLFLN40X). Illumination was provided by a 520-nm laser, and the resulting images were captured using a CMOS image sensor (Basler acA2440-75um). The phase distribution of resolution targets were generated computationally by directly assigning constant phase values from \SI{0.5}{\radian} to \SI{1.5}{\radian} and unit amplitude to specific areas. For these samples, PCM intensity images are simulated by applying the PCM model in \eqref{eqn:pcm} on the phase images. Therefore, the model used to solve the inverse problem is exact.

\begin{table}[tb]
    \centering
    \caption{Root-mean-square (RMS) error (in rad) for simulated PCM.\footnote{UNN = untrained neural network. R-H, R-M, R-L = regularization high, mid, and low. The restoration process was carried out 5 times for each individual sample and the average RMS error is shown.}}
    \begin{ruledtabular}
    \begin{tabular}{lcccc}
         & Beads & \makecell{Separated\\Cells} & \makecell{Grouped\\Cells} & \makecell{Resolution\\Targets} \\
        \colrule
        UNN & \textbf{0.040} & 0.199 & \textbf{0.240} & \textbf{0.243}\\
        R-H & 0.043 & 0.222 & 0.289 & 0.369 \\
        R-M & 0.043 & 0.200 & 0.265 & 0.449 \\
        R-L & 0.070 & \textbf{0.191} & 0.284 & 0.660\\
    \end{tabular}
    \end{ruledtabular}
    \label{tab:rms_table}
\end{table}

Table \ref{tab:rms_table} lists the root-mean-square (RMS) errors of the restored phase images compared to the ground truth images. On the whole, our proposed method outperforms the best results of the three regularization groups in most cases. Even in the worst case -- the separated cell group -- our method still provides comparable restoration quality. Each of the regularization groups can perform fairly well in restoring some of the sample groups, but on no occasion in restoring all of them. This result demonstrates the generalizability of the self-adaptive UNN-based approach and highlights the lack of generalizability as an issue with the manual-regularization-based approach.

\begin{figure}[tb]
    \centering
    \includegraphics[width=\linewidth]{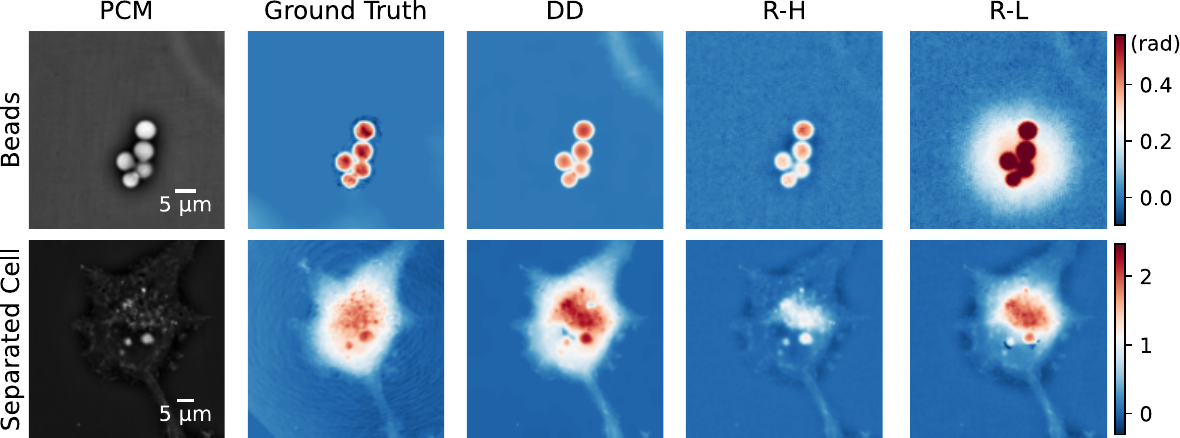}
    \caption{Comparison between phase retrieval by UNN and by regularization. PCM images, ground truth phase images, and phase restorations of a beads sample and a separated cell sample are exhibited.}
    \label{fig:comparison}
\end{figure}

A similar restoration process was carried out using experimental PCM images. To acquire PCM images, we used the same microscope and image sensor as above but the microscope was equipped with a 525-nm LED, a condenser annulus, and a negative phase contrast objective (UPlanFLN 40x/0.75NHPh2). Unlike simulated ones, experimental PCM images are affected by model inaccuracies, experimental errors, and other factors. In Table \ref{tab:table_exp}, we compared the restoration accuracy of experimental PCM images of two kinds of samples: 1) 19 images of microbeads and 2) 6 images of separated cells. Their phases were acquired using DH in the same way as described above, except that the illumination wavelength for the separated cells was \SI{532}{\nano\meter}. As anticipated, UNN again outperformed the regularization-based method, confirming its effectiveness in an actual optical setup with experimental errors. As a specific example, Fig.\ \ref{fig:comparison} illustrates restoration results of one sample from each group. It can be seen that, while stronger regularization successfully restores the beads sample, it fails with the cell sample; conversely, weaker regularization only succeeds with the cell sample. This can be explained by noticing that in the beads sample, the background area occupies a larger portion, making the image sparser, where strong regularization is appropriate. However, for the cell sample, this regularization strength imposes excessive constraints, preventing effective restoration.

\begin{table}[tb]
    \caption{RMS error (in rad) for experimental PCM.}
    \begin{ruledtabular}
    \begin{tabular}{lcccc}
         & Beads & \makecell{Separated\\Cells} \\
        \colrule
        UNN & \textbf{0.044} & \textbf{0.172} \\
        R-H & 0.045 & 0.327 \\
        R-M & 0.045 & 0.290 \\
        R-L & 0.069 & 0.199 \\
    \end{tabular}
    \end{ruledtabular}
    \label{tab:table_exp}
\end{table}

\begin{figure}[tb]
    \centering
    \includegraphics[width=\linewidth]{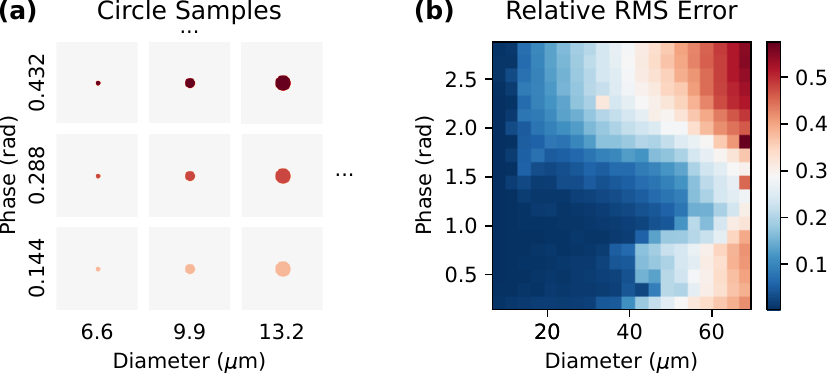}
    \caption{(a) Artificial circle samples of different diameters and phase values. (b) Relative RMS error averaged over 5 restorations with each circle sample.}
    \label{fig:heatmap}
\end{figure}

Limited by its imaging principle, PCM does not preserve low spatial-frequency information \cite{nguyen2017halo}. As a result, it is difficult to recover this part of information by any phase retrieval algorithm without prior knowledge of the object. Due to this reason, samples of too large size cannot be retrieved efficiently, regardless of the method. To estimate the range of recoverable samples by our method, we have performed simulation over artificial circle samples of a series of diameters and phase values (Fig.\ \ref{fig:heatmap}(a)). For each sample the phase retrieval process is repeated 5 times, and the average relative RMS error is summarized in Fig.\ \ref{fig:heatmap}(b). It is found that for samples with phase value $\lesssim\SI{1.5}{\radian}$, the maximum diameter allowing for a decent restoration is about \SI{40}{\micro\meter}. For smaller samples whose diameters $\lesssim \SI{12}{\micro\meter}$, phase restoration is possible up to a phase value as large as \SI{2.88}{\radian}, which is the upper bound limited by the phase delay and transmittance of the phase ring \cite{kurata2024single}.

\section{Discussion}
Although UNN restoration performed better on both the beads and separated cells, it should be noted that hole artifacts occasionally appear inside the sample, as illustrated by the separated cell sample in Fig.~\ref{fig:comparison}. This kind of artifact could be residues from the initialization of the deep decoder weights that were not eliminated during optimization. Besides gradient descent with an Adam optimizer, as mentioned earlier, we also tried projected gradient descent, but it was unable to remove the artifacts either. A temporary solution could involve a pretraining process, where a segmentation image $\mathcal{S}\qty(\bm{g})$ of the raw PCM images serves as an initial guess for the phase distribution $\bm{\theta}$. The deep decoder would then be trained for several hundred epochs to approximate this $\mathcal{S}\qty(\bm{g})$ (See Appendix \ref{sec:holes}). It would be useful to develop an optimization process specifically tailored to the compressive propagation setup. Additionally, further analysis and consideration of the convergence of our algorithm should be conducted based on previous work \cite{jagatap2019algorithmic}.

\section{Conclusion}
In this study, we have implemented a phase retrieval algorithm using a deep decoder image prior to retrieve phase from PCM intensity images, achieving quantitative phase imaging. Our proposed approach eliminated the need for empirical hyperparameter tuning, which is required in the earlier regularization-based approach. We have compared the restoration accuracy of our method with that of the regularization-based method for both simulated and experimental PCM images. Our approach demonstrated improved accuracy and robustness across various samples, illustrating the deep decoder's capability as a self-learned regularizer. The success of our approach indicates that the deep-prior-based phase retrieval algorithm is compatible with the compressive propagation setup. Our study pushes the boundaries of QPI by utilizing off-the-shelf PCM, which is widely employed in biological and medical fields.


\appendix

\section{Phase Samples}\label{sec:samples}
\begin{figure}[htbp]
    \centering
    \includegraphics[width=\linewidth]{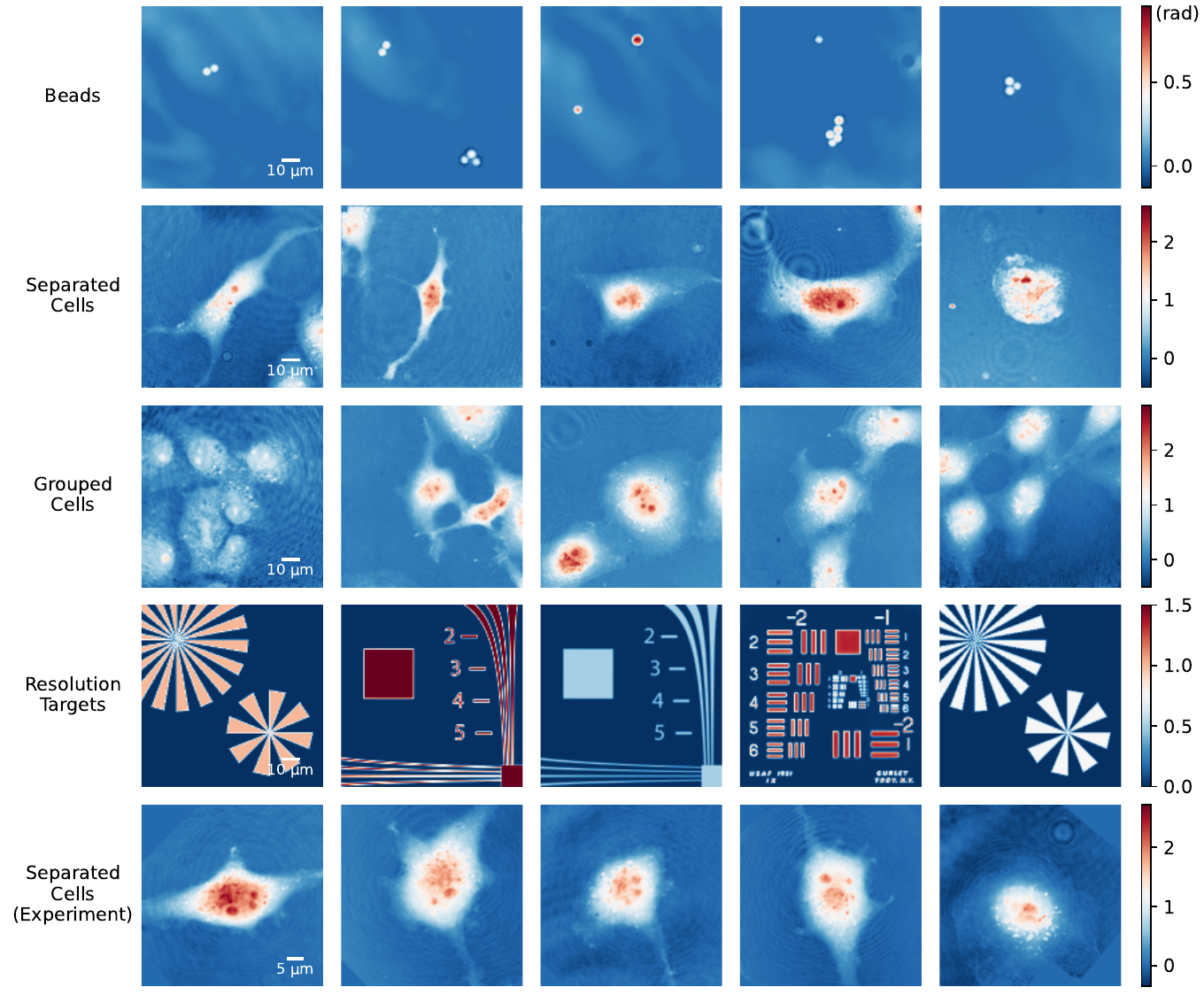}
    \caption{A preview of phase samples. ``Beads" group is used for recovery from both simulated and experimental PCM images; ``separated cells", ``grouped cells", and ``resolution targets" for simulated ones only, ``separated cells (experiment)" for experimental ones only.}
    \label{fig:preview}
\end{figure}

Here we show a preview of phase samples used in this work (Fig.\ \ref{fig:preview}). Resolution target samples are phase-only samples which are generated computationally. We created three such targets as shown in Fig.\ \ref{fig:preview}, and each of them is assigned to five phase values $\qty{0.5, 0.75, 1, 1.25, 1.5} \si{\radian}$, rendering 15 sample images in total. The other samples are recorded by a quantitative phase imaging platform based on off-axis digital holography.

\section{Hole Artifacts}\label{sec:holes}
\begin{figure}[htbp]
    \centering
    \includegraphics[width=0.95\linewidth]{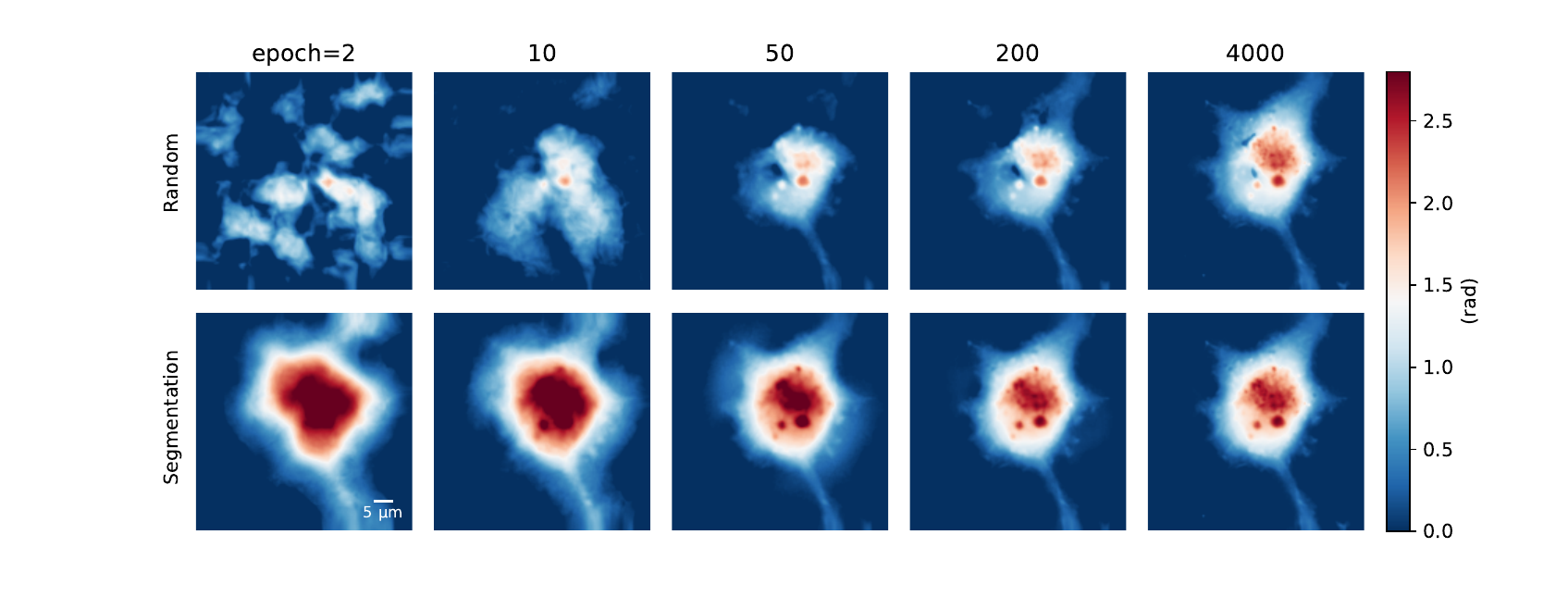}
    \caption{Visualization of estimated phase images during training process with (bottom) and without (top) pre-training.}
    \label{fig:holes}
\end{figure}

By inspecting the updating of estimated phase, it is apparent that the hole artifacts is associated to the initial phase distribution determined by random initialization of deep decoder weights (Fig.\ \ref{fig:holes} top). A temporary solution is to use an initial guess of phase. In Fig.\ \ref{fig:holes} bottom, a segmentation image $\mathcal{S}\qty(\bm{g})$ of the raw PCM image, created through edge detecting and Gaussian filtering, was used as an initial guess for $\bm{\theta}$. The deep decoder was first trained for 500 epochs to approximate this $\mathcal{S}\qty(g)$. The hole artifacts can be somewhat suppressed by this pre-training. The problem is that, when the sample is complex and noisy, simple segmentation algorithm usually cannot work well.

\section{Computational Resources}
All numerical experiments were conducted on a desktop computer equipped with an Intel Core i9-13900K CPU and an NVIDIA TITAN V GPU. The deep prior based phase retrieval algorithm, implemented in PyTorch with CUDA support, primarily leveraged the GPU and took approximately 70 seconds to restore one image.

\nocite{*}

\bibliography{apssamp}

\end{document}